\documentclass[pdflatex,sn-mathphys-num]{sn-jnl}% Math and Physical Sciences Numbered Reference Style
\usepackage{graphicx}%
\usepackage{multirow}%
\usepackage{amsmath,amssymb,amsfonts}%
\usepackage{amsthm}%
\usepackage{mathrsfs}%
\usepackage[title]{appendix}%
\usepackage{xcolor}%
\usepackage{textcomp}%
\usepackage{manyfoot}%
\usepackage{booktabs}%
\usepackage{algorithm}%
\usepackage{algorithmicx}%
\usepackage{algpseudocode}%
\usepackage{listings}%
\usepackage{array}

\usepackage{pifont}

\newcommand{\upon}[1]{\State \textbf{upon} #1\textbf{ do}}

\DeclareUnicodeCharacter{00A0}{ }
\theoremstyle{thmstyleone}%
\theoremstyle{thmstyletwo}%

\theoremstyle{thmstylethree}%

\begin{document}

\title[Article Title]{T-RBFT: A Scalable and Efficient Byzantine Consensus Based on Trusted Execution Environment for Consortium Blockchain}

%%=============================================================%%
%% GivenName	-> \fnm{Joergen W.}
%% Particle	-> \spfx{van der} -> surname prefix
%% FamilyName	-> \sur{Ploeg}
%% Suffix	-> \sfx{IV}
%% \author*[1,2]{\fnm{Joergen W.} \spfx{van der} \sur{Ploeg} 
%%  \sfx{IV}}\email{iauthor@gmail.com}
%%=============================================================%%

\author[1]{\fnm{Wen} \sur{Gao}}

\author[1,2]{\fnm{Xinhong} \sur{ Hei}}

\author*[1,2]{\fnm{Yichuan} \sur{Wang}}\email{chuan@xaut.edu.cn}

\affil[1]{
	\orgdiv{School of Computer Science and Engineering}, 
	\orgname{Xi'an University of Technology}, 
	\orgaddress{\city{Xi'an}, \postcode{710048}, \country{China}}
}

\affil[2]{\orgname{Shaanxi Key Laboratory for Network Computing and Security Technology}, \orgaddress{\city{Xi'an}, \postcode{710048}, \country{China}}
}

\abstract{With the continuous expansion of blockchain application scenarios, consortium chains have raised higher performance and security requirements for consensus mechanisms.  Unlike public blockchains, consortium chains typically implement an admission mechanism that restricts participation to trusted entities, ensuring that most replicas are honest and the number of faulty nodes remains small under normal circumstances. In such settings, conventional Byzantine Fault Tolerant (BFT) protocols, which are designed for worst-case adversarial scenarios, incur excessive message exchanges and computational overhead, thereby limiting performance and scalability. To address this issue, this paper proposes T-RBFT, a two-layer consensus mechanism inspired by network sharding and enhanced by the trusted execution environment (TEE).   In T-RBFT, consensus nodes are first dynamically grouped based on their runtime characteristics.  Then, inter-group consensus is achieved through a TEE-assisted BFT protocol, while each group internally reaches agreement using an improved Raft-based mechanism. Experimental evaluation shows that T-RBFT reduces communication overhead and latency, and achieves higher throughput compared to existing two-layer consensus protocols, providing a scalable and communication-efficient consensus protocol for permissioned blockchain networks.}

\keywords{Two-layer consensus protocols, Byzantine fault-tolerance, Trusted execution environment, Consortium blockchain}

%%\pacs[JEL Classification]{D8, H51}

%%\pacs[MSC Classification]{35A01, 65L10, 65L12, 65L20, 65L70}

\maketitle

\section{Introduction}
Blockchain is a distributed database technology realized through a variety of technologies such as public key cryptography algorithms, hash algorithms, consensus mechanisms, and distributed storage technologies. Among them, the consensus mechanism focuses on solving the consistency problem of the blockchain, aiming to ensure the consistency of the data copies maintained by all nodes and to avoid the occurrence of data inconsistency and information asymmetry problems.

As one of the important branches of blockchain technology, the consensus mechanism of the consortium blockchain needs to take into account the efficiency, safety and scalability, especially in the large-scale network environment, it still need to maintain high throughput and low latency. Currently available consensus mechanisms are not yet able to meet these needs at the same time. For example, IBM's Hyperledger Fabric blockchain \cite{1} previously supported PBFT \cite{2} in earlier versions, but it was eventually phased out in favor of other mechanisms like Raft and Kafka. This decision was primarily driven by the scalability challenges of PBFT, where communication overhead grows quadratically with the number of nodes, leading to a performance bottleneck. Therefore, a large number of research works in recent years have improved the performance of Byzantine fault-tolerant (BFT) algorithm by optimizing the consensus structure \cite{3,4,5,6,7}, electing the committee \cite{8,9,10}, choosing suitable signature algorithms or underlying communication modes \cite{11,12,13,14} and calculating node reputation values to elect high reputation nodes as consensus participants \cite{30,31} among other ideas. However, only improving the BFT algorithm cannot fundamentally solve the problems of the BFT algorithm. When the number of nodes increases, the three-phase submission process will still cause its performance to decrease significantly.

Another idea is to combine the BFT consensus algorithm with trusted hardware, using the security of  hardware  to reduce the number of replicas or communication stages in the BFT protocol.  For example, Minbft \cite{16} implements a monotonically increasing counter function in trusted platform module (TPM), which uniquely binding the value $i$ of a certain counter to certain messages, so that the node cannot perform different $i$-th operations, thereby causing nodes to degenerate from Byzantine nodes to non-Byzantine nodes, and reducing the number of communication stages from 3 to 2. Under normal circumstances,
FastBFT \cite{18} proposes a tree structured communication mode and combines hardware-based TEE with lightweight secret sharing, further improving the scalability of the BFT algorithm. However, since malicious nodes can deliberately cause member replacement of tree communication, the operation of FastBFT relies on a relatively stable cluster environment. NefSBFT \cite{33}  has designed a propagation technique that is executed in the improved FastBFT to achievetransaction ordering and block verification, solving the problem of low message delivery rate in FastBFT.
Although these solutions have been optimized in terms of communication complexity and number of nodes, most of them rely on a global serial consensus process and therefore suffer from performance bottlenecks and limited scalability when faced with high concurrency or large-scale node deployment \cite{17,19}.

In order to effectively break through the performance bottleneck that restricts BFT applications, there have been two-layer consensus schemes in recent years to integrate the Crash Fault Tolerance (CFT) algorithm with the BFT algorithm, avoiding the low availability of  consortium blockchain due to too long consensus time. For example,  Huang et al. \cite{xinsan} proposed RepChain which combines a synchronous Byzantine fault tolerance and Raft protocols to complete consensus. Li et al. \cite{20} proposed a two-layer consensus mechanism called R-PBFT  with supervised nodes, which solves the performance bottleneck problem of consortium blockchain in a large-scale node network environment. Zhao et al. \cite{32} proposed a new two-layer consensus protocol  based on the primary-secondary consortium chain called WRBFT. WRBFT uses weighted Raft to complete the intra-group consensus, while the inter-group consensus uses  PBFT algorithm based on Boneh-Lynn-Shacham (BLS) signatures and the verifiable random function.   Yuan et al. \cite{jia2} proposed a two-layer consensus algorithm called  DLCA\_R\_P based on the improved RAFT consensus algorithm and PBFT consensus algorithm. DLCA\_R\_P has further improved consensus efficiency by using a supervision mechanism and reputation mechanism. However, these schemes have significantly reduced the number of Byzantine nodes can be tolerated, resulting in safety that needs to be improved.

Since the consensus mechanism of  consortium blockchain needs to take into account efficiency, safety and scalability, this paper proposes T-RBFT (TEE-based Raft cluster Byzantine Fault Tolerance), a two-layer consensus mechanism based on trusted execution environments. adopts the concept of network sharding to partition consensus nodes into groups, transforming the traditional global BFT consensus into a more scalable inter-group and intra-group consensus model. Within each group, an improved Raft algorithm is employed to achieve fast and efficient local consensus, while a TEE-based BFT protocol is used among the leaders of each group to ensure Byzantine fault tolerance at the global level.

Our experiments show that T-RBFT outperforms the most advanced two-layer consensus protocols \cite{20,32} in terms of consensus latency, throughput, and fault tolerance.

To sum up, the contributions of this paper can be summarized as follows:

$\bullet$ 	We propose a node grouping method based on consistent hashing and dynamic weights to improve the rationality and maintainability of grouping construction.

$\bullet$ 	With the help of the trusted execution environment, we  propose a two-layer consensus mechanism suitable for  consortium blockchain. 
Experimental results show that even with trusted component access overhead, the consensus mechanism in this paper can have better throughput than existing two-layer consensus protocols, and have smaller latency in networks with non-negligible communication delays.

$\bullet$ 	We have designed a strengthening mechanism for the intra-group consensus from three aspects: log replication, leader election and committed confirmation, which enables intra-group consensus to have byzantine fault tolerance on the basis of the same understandability as Raft.

The rest of the paper is arranged as follows. Section 2 introduces the related knowledge of technology used in this paper.  In  Section 3 and 4, we propose T-RBFT consensus mechanism and its detailed design.  After that, we provide the complete proof of correctness of T-RBFT in  Section 5 and give the evaluation results in Section 6. Finally, in Section 7, we summarize this paper and look forward to the future work.  

\section{Related Work}
In this section, we discussed the bottlenecks of PBFT and Raft consensus algorithms, as well as the TEE and USIG services that will be used in our experimental scheme.
\subsection{Practical Byzantine Fault Tolerance}
The PBFT algorithm was proposed by Castro and  is widely used in scenarios that require high fault tolerance. In PBFT, if $n=3f+1$ , then a system with $n$  nodes can tolerate   $f$ faulty nodes. During the execution of the PBFT protocol, one node is referred to as the primary, while the other nodes are called replicas, and all nodes are connected to the client of the service \cite{21}. The protocol consists of three phases: PRE-PREPARE, the master node sends the proposal to all replicas; PREPARE,  the replicas check the proposal and broadcast votes with other replicas; COMMIT,  a replica that has received at least $n=2f$  prepare completion messages sends commit messages to all other replicas. Once a replica has collected $n=2f+1$  commit messages, it executes the request and returns the message to the client. Due to the protocol broadcasting messages twice in the later two stages, its complexity is $O(N^{2})$. When there are too many nodes in the network, its performance will be significantly reduced.

\subsection{Raft consensus algorithm and its bottlenecks}
The Raft protocol \cite{34} is a fault-tolerant protocol with strong leadership capabilities. The communication complexity is $O(N)$. 
In the Raft, nodes have three possible states: follower, candidate, and leader. The Raft mechanism divides time into several time periods of arbitrary length, and each time period is called a term. During each term, there is only one leader. 

In terms of fault tolerance, Raft can tolerate no more than 50\% of downtime failures, but it does not have Byzantine fault tolerance. Although Raft is a non-Byzantine fault-tolerant consensus algorithm with high performance, if it is applied to an environment with Byzantine nodes, the current restrictions added by Raft to ensure safety will not be enough to continue to ensure safety, which are mainly reflected in the following three aspects \cite{22}:

1) Log replication

\ding{172} Assume that the leader node is a Byzantine node: the transaction information may be tampered with by the Byzantine node, causing the follower to receive incorrect transaction information; or the leader will send different block information to different followers, leading to blockchain inconsistency.

\ding{173} Assume that the follower node is a Byzantine node: the follower node may tamper with the received block content, add it to its own blockchain, and reply to the leader that it has successfully uploaded the block to the chain, which may result in less than half of the nodes having the same blockchain. At this time, the node may actively propose to run for the leader and overwrite the original block.

Therefore, if Byzantine nodes exist, Raft's original "Leader's log list must be the most complete log list and log entries will only be copied unidirectionally from leader to follower" mechanism can no longer guarantee safety. Different nodes maintain inconsistent log lists, resulting in the final state of the state machine no longer being consistent.

2) Leader election

Assuming that candidate is a Byzantine node, the logical timestamp in the RequestVote message can be forged by the Byzantine Candidate to be "newer" than the actual value, thereby obtaining votes from other nodes. Therefore, after the Byzantine candidate is elected as leader, it can overwrite the committed log entries in the log lists of other nodes by copying  logs, which cannot guarantee safety.

3) Consensus confirmation

\ding{172} Assume that the leader node is a Byzantine node: 
it may return to the client that consensus has been reached without more than half of its followers responding.

\ding{173} Assume that the follower node is a Byzantine node: It can claim to have added a block to its own blockchain without actually adding it, which in extreme cases affects the cluster's ability to reach consensus and cannot guarantee safety in a Byzantine fault-tolerant environment.

Therefore, Raft's original method of judging whether a log entry is committed can no longer guarantee safety in a Byzantine fault-tolerant environment.

\subsection{Hardware security mechanism}
Trusted execution environment (TEE) \cite{23} is a privacy computing technology based on the security architecture of hardware characteristics and system software layer (operating system or virtualization system). It can build multiple secure computing environments called trusted execution environments in the computer. Each TEE is capable of running universal algorithm logic internally and implementing data confidentiality calculations. The security model of TEE technology ensures that any external attacker cannot steal the algorithms or data running inside of the TEE, nor can it maliciously control the execution of the algorithms within the TEE, fully ensuring data privacy and program integrity. TEEs are currently popular on mobile platforms \cite{24}, while TEEs such as Intel's SGX \cite{25} and TPM \cite{26} are being deployed on PCs and servers.

\subsection{USIG service}
USIG is a unique sequential identifier generator that assigns counter values to a message (i.e. byte arrays) and signs it. To prevent ambiguity, we require that this service must be implemented in a tamper-resistant component. USIG has the following three properties:

\ding{172} Uniqueness: The same identifier will not be assigned to two different messages.

\ding{173} Monotonicity: No lower identifier than the previous one will be assigned.

\ding{174} Sequential: No identifier will be assigned to a successor who is not the previous identifier.

USIG service exposes two functions through its  interface. The first, $\textsf{createUI}(m)$, generates a USIG certificate that encapsulates a unique identifier ($UI$) tied to the message $m$. This identifier is obtained from a strictly increasing monotonic counter maintained  within the tamper-proof component, which guarantees that each invocation results in a distinct $UI$. The USIG certificate thus serves as evidence that the message was processed by a trusted instance and assigned a unique, ordered identifier at the time of creation; the second, $\textsf{checkUI}(PK,UI,m)$, checks  whether the provided $UI$ is valid for the message $m$ using the public key $PK$, ensuring that the USIG certificate is consistent with the message content and was issued by the correct USIG instance.

This service can be implemented using a Hash-based Message Authentication Code (HMAC), where the USIG certificate is generated via  HMAC using a shared secret key. This requires that both the $\textsf{createUI}$ and $\textsf{checkUI}$ functions be executed entirely within the tamper-proof component, such as a TEE, to protect the secrecy of the  key and maintain the service's integrity and authenticity.

\section{The T-RBFT Mechanism}
In this section, we give an overview of T-RBFT before providing a detailed specification in Section 4.

\subsection{Model Assumptions}
We focus on consortium blockchain scenarios, where all consensus nodes operate in a reliable and well-managed network environment, and therefore adopt the partially synchronous network model as our basic assumption. Specifically, the system contains $N$ consensus nodes, which are divided into $k$ groups to provide Byzantine fault-tolerant consensus services for the blockchain. The network may exhibit arbitrary delays for an unknown period of time, but eventually becomes synchronous, where message transmission delays are bounded.

In a consortium blockchain environment, where access control, identity authentication, and a baseline level of trust are already established, absolute decentralization is not required-although malicious behavior must still be prevented. Therefore, it is feasible to designate a "scheduler" or "organizer" to handle the node grouping process. We assume that this task is managed by a trusted consortium administrator or a scheduling committee, and that the grouping process is publicly auditable to ensure fairness and transparency.

We assume that each node is configured with a trusted execution environment. TEEs can use remote attestations to authenticate each other and establish secure communication channels between them.

In addition, attackers may tamper with messages through the network, so a digital signature or HMAC needs to be included in the message. We assume that nodes and clients know each other's keys in order to check these signatures/HMAC. Furthermore, we think that the hash function is conflict resistant, meaning that signatures cannot be forged.

\subsection{Overview of T-RBFT Consensus Mechanism}
In order to provide a consensus mechanism with high security, low latency, and high throughput, we use TEE to propose a two-layer consensus mechanism called T-RBFT  for consortium blockchain. T-RBFT includes two phases: construction of node grouping and consensus process, which would be described in Section 4 in detail.

\begin{figure}[htbp]\centering
	\includegraphics[width=0.7\columnwidth]{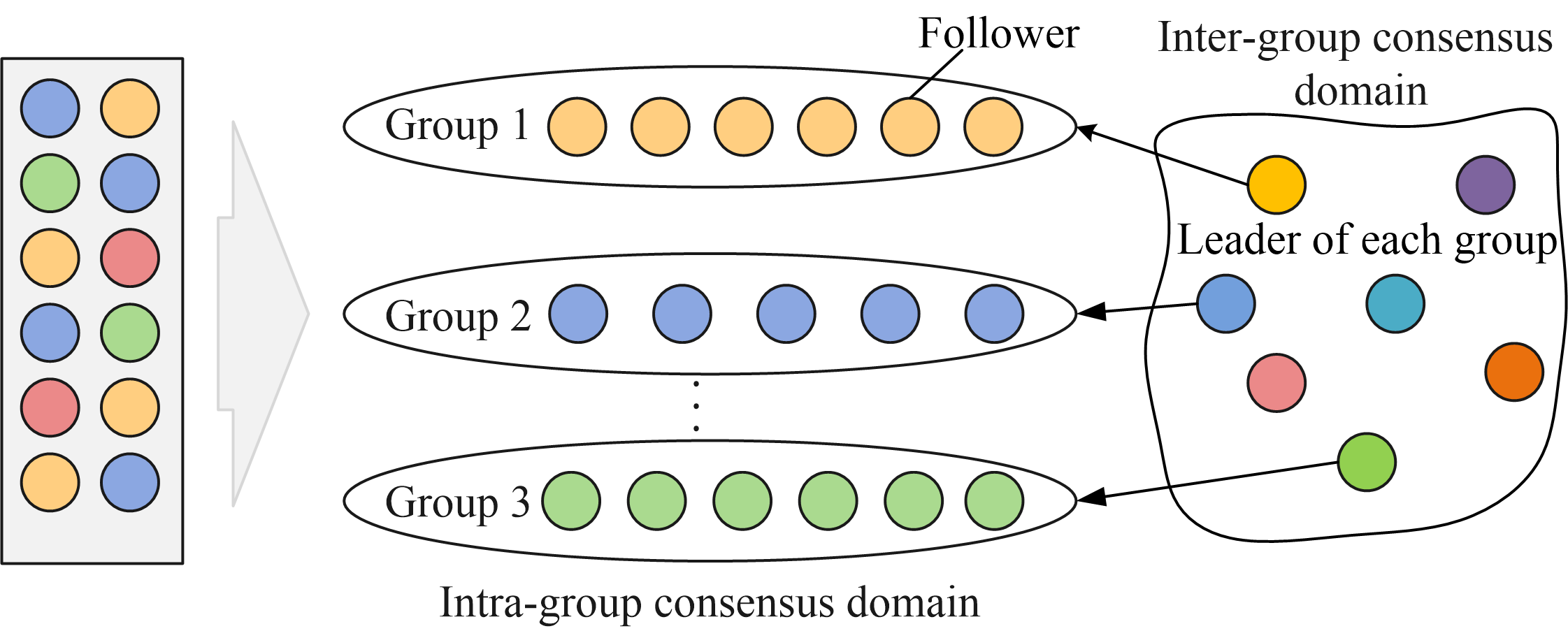}
	\caption{Consensus  nodes composition diagram for T-RBFT.\label{1}}
\end{figure}

Construction of node grouping includes initial grouping and dynamic weight grouping. Initial grouping is to divide the initial consensus nodes of consortium blockchain into multiple groups according to the grouping strategy. Dynamic weight grouping is suitable for situations where nodes dynamically join or exit during the operation of  blockchain network. In order to ensure the balance of the grouping,  we assign the node to corresponding group according to the grouping weight. 
The nodes composition diagram  is shown in Fig.~\ref{1}. Each group elects a leader equipped with  TEE, and the remaining nodes serve as followers.

Then, the system performs consensus on the client's request messages. The consensus process is divided into two stages: inter-group consensus and intra-group consensus. 
The message exchange pattern for inter-group consensus is similar to PBFT. However,the difference is that each of leaders implements the USIG service, the TEE-based USIG service can not only reduce the minimum number of participating inter-group consensus nodes from $3f + 1$ to $2f + 1$, but also make the entire inter-group consensus process only have two communication steps instead of three communication steps like PBFT.
%Then, T-RBFT execute ?inter-group consensus? among all leaders relying on the trusted subsystem. 
After the inter-group consensus is completed, the intra-group consensus stage is entered. The message exchange pattern for intra-group consensus is similar to Raft. The difference is that we use  TEE-based USIG service to design strengthening mechanisms from three aspects: log replication, leader election and consensus confirmation, so that the intra-group consensus has byzantine fault tolerance.

\section{T-RBFT: Detailed Design}
This section introduces  our node grouping strategy and describes in detail the consensus process of the T-RBFT under normal case and the view changing operation when the primary is faulty.

\subsection{Construction of node grouping}
Firstly, T-RBFT adopts the idea of network sharding to grouping blockchain nodes. Existing network sharding schemes mainly includes protocol-based sharding \cite{28} and geographically-based sharding \cite{29}. However, these two methods can easily lead to uneven sharding and ignore the performance differences of nodes, which cannot ensure node balance within sharding. Therefore, this paper proposes a new node grouping method. The specific process is as follows.

\subsubsection{Initial grouping} 
In the initial grouping phase, we adopt the grouping method inspired by \cite{20}. Specifically, we utilize the design concept of the consistent hashing algorithm and leverage its balanced, decentralized, and monotonic properties to divide the nodes into multiple groups. The process is as follows.

1) InitHashRing ($k$) // Initialize the hash ring according to the number of groups $k$, and determine the position of each group ID. 

2) Nodes grouping

Calculate Hash (nodeID + ip + randomStr), Map nodes to rings based on results. 

ClockwiseSearch ( ) // Find the group ID clockwise according to the node position, and map the node to the corresponding different groups. 

3) Rationality verification 

Verify that each group contains at least $n \geq 3$ nodes. If the condition is met, proceed to Step~4; otherwise, return to Step~2 to adjust the grouping.

4) Grouping completed

\begin{figure}[htbp]\centering
	\includegraphics[width=0.7\columnwidth]{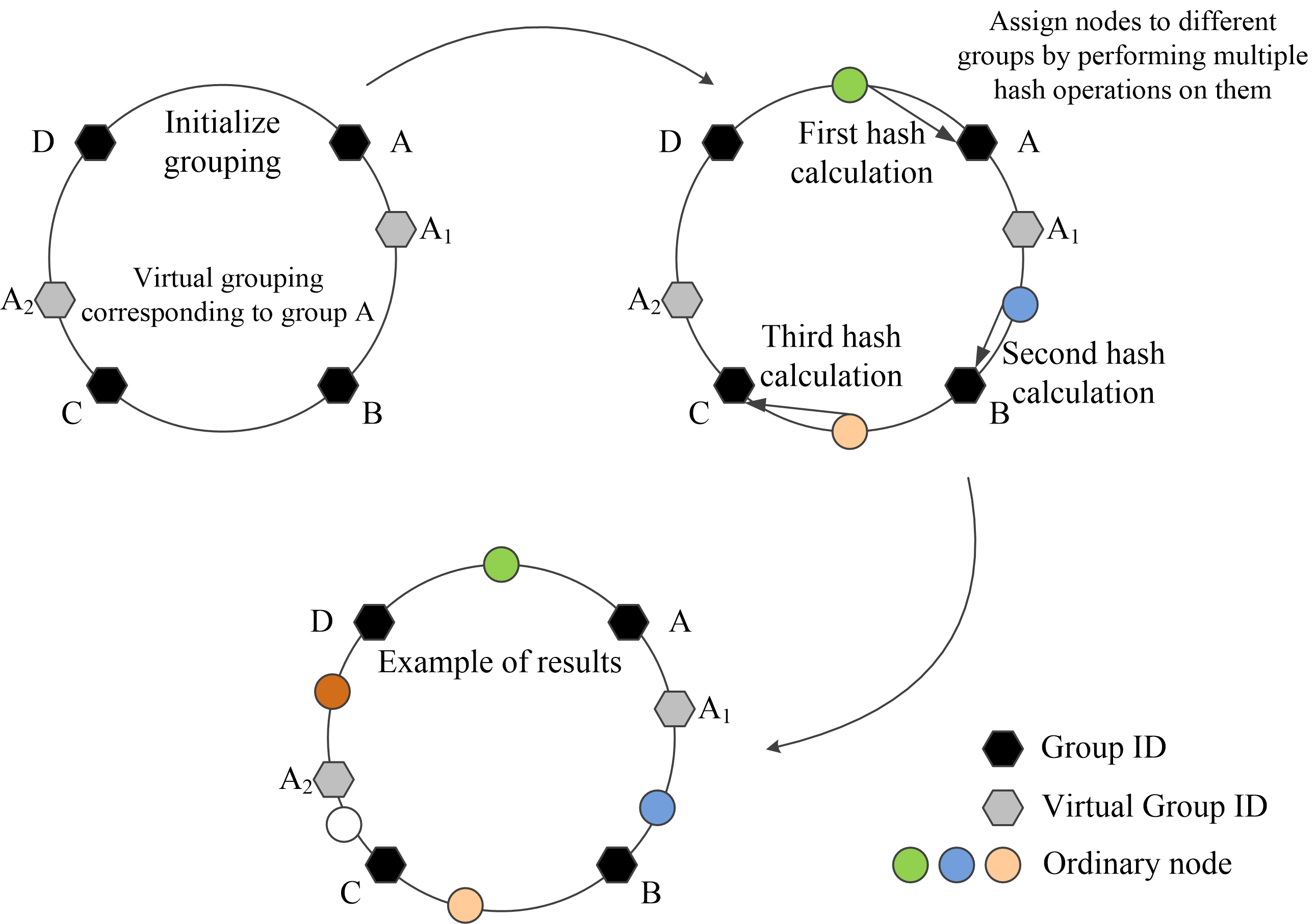}
	\caption{Grouping strategy using consistency hash algorithm.\label{2}}
\end{figure}

The grouping strategy using the consistent hashing algorithm is shown in Fig.~\ref{2}. The result of the consistent hash calculation is a 32-bit unsigned integer (i.e., a value of type uint32), which also has the collision resistance of hash. Based on the result, the hash value of the node mapping can be distributed on a circle from 0 to $2^{32}$. However, when the number of groups is small, it is easy to cause data skewing problems due to uneven grouping, that is, some groups carry the majority of nodes, resulting in load imbalance. Therefore, the algorithm introduces the concept of virtual nodes in the consistent hashing algorithm,  corresponding actual grouping to multiple virtual grouping to expand the number of groups and make the initial grouping more uniform \cite{15}.

\subsubsection{Dynamic weight grouping}
Assume that the number of existing nodes in the network is $N$, the number of groups is $k$, the $j$-th group is represented by $G_{j}$. To maintain balance among groups when new nodes join,  we define a composite score $S_{j}$ for each group, which integrates both its credit value and load.  Specifically, the credit value is positively correlated with the score, as a higher credit indicates stronger consensus stability, while the load is negatively correlated, as high resource usage suggests limited capacity for additional nodes.

1) Overall credit value of grouping

The credit value of each group is mainly affected by two parameters: consensus success rate and transaction impact rate, which can be used to calculate the credit value of each group.

The consensus success rate $\gamma$ refers to the situation where $j$-th group $G_{j}$ successfully reaches a consensus  and its calculation formula is:
\begin{equation}
\gamma=\alpha\sqrt{\frac{t}{n}}
\label{equation2}
\end{equation}

Here, $t$ represents the number of times group $G_{j}$ in the blockchain has successfully reached consensus, $\alpha$ is a adjustment constant, and $n$ represents the total number of times group $G_{j}$ has participated in consensus.

The transaction impact rate $f(x)$ is used to indicate the degree of impact of a transaction on  group  $G_{j}$ and its calculation formula is:
\begin{equation}
f(x) =
\begin{cases}
(\frac{x}{M_{0}})^{\frac{1}{2}}   &  x< M_{0} \\
1
&  x\geq M_{0}
\end{cases} 
\label{equation3}   
\end{equation}

Here, $M_{0}$ represents the limit value of the importance level of  transaction, $x$ represents the importance level of transaction, and the value of $f(x)$ is positively correlated with the value of $x$.

Therefore, the formula for credit values $C$ of group $G_{j}$ is:
\begin{equation}
C(G_{j})=\gamma\frac{\sum\nolimits_{x=1}^n E f(x)}{\sum\nolimits_{x=1}^n  f(x)}+C_{init}
\label{equation5}
\end{equation}

Here, $C_{init}$ represents the initial credit values of  group $G_{j}$, and $E$ represents the credit impact factor.

2) Group load

The load of node $n_{i}$, which belongs to  group $G_{j}$, is defined as a weighted combination of resource utilization metrics, including CPU usage $L_{cpu}(n_{i})$, memory usage $L_{mem}(n_{i})$, and network utilization $L_{net}(n_{i})$.
So, the load $L(n_{i})$ of node $n_{i}$ is computed as:
\begin{equation}
L(n_{i})=\beta.L_{cpu}(n_{i})+\gamma.L_{mem}(n_{i})+\delta.L_{net}(n_{i})
\label{equation1}
\end{equation}

Here, $\beta+\gamma+\delta=1$, which defines the relative importance of each dimension in the node load calculation. Based on this, the average load of   group $G_{j}$ (i.e., the group load) is computed as:

\begin{equation}
\bar{L}(G_{j}) = \frac{1}{|G_j|} \sum_{n \in G_j} L(n)
\label{equation4}
\end{equation}

Therefore, the composite score of group $G_{j}$ is  computed as:
\begin{equation}
S_{j} =w_{1}\cdot C(G_{j}) -w_{2}\cdot \bar{L}(G_{j})
\label{equation7}
\end{equation}

Here, $w_{1}$ and $w_{2}$  denote the weights assigned to the group credit and the group load, respectively. 

When a new node joins the system, it is assigned to the corresponding groups with the highest score, which represents an optimal trade-off between group trustworthiness and current resource pressure. As system states evolve (e.g., changes in group credit or load), the composite scores are periodically recalculated to maintain the timeliness and effectiveness of the grouping strategy.

In addition, in order to better control the resource consumption of the grouping process, we use the regrouping upper limit threshold and the regrouping lower limit threshold parameters. When a node joins or exits the blockchain network, the number of nodes in each group will be checked. If the number of nodes in a certain group is greater than the upper regrouping threshold or less than the lower regrouping threshold, the blockchain will be re-grouped according to the initial grouping method.

\subsection{Consensus process of T-RBFT}
\begin{figure}[htbp]\centering
	\includegraphics[width=0.8\columnwidth]{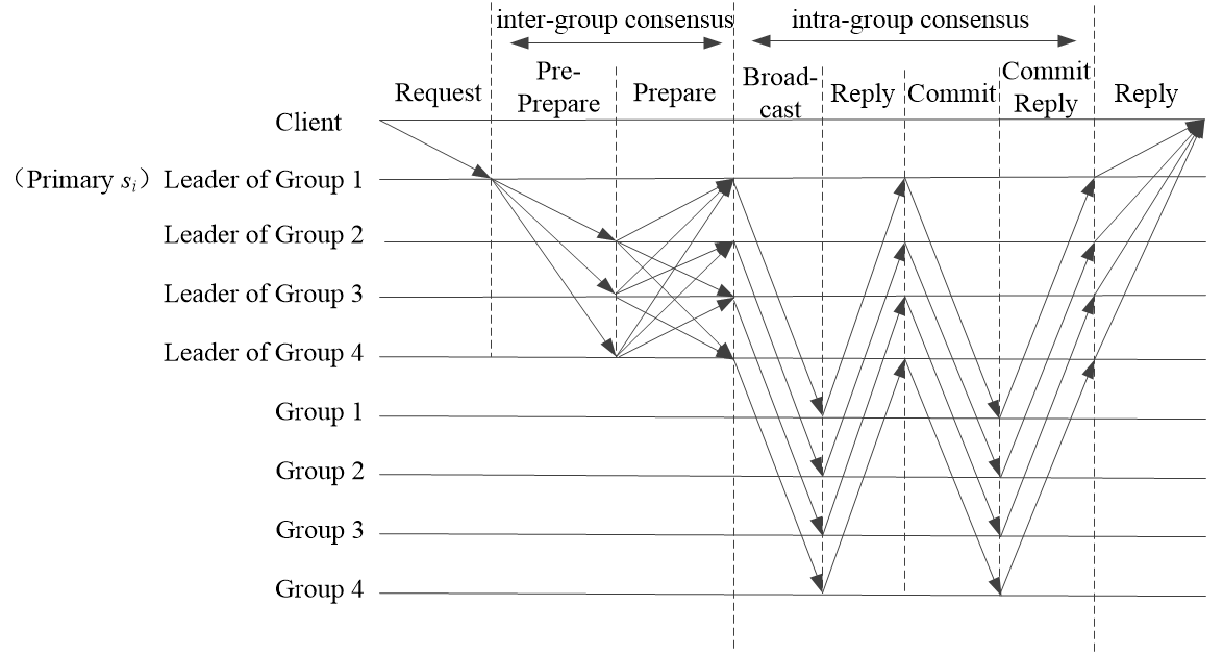}
	\caption{Consensus flow of T-RBFT algorithm.\label{3}}
\end{figure}

The primary packages several pieces of  received client's request information into a block for consensus. The consensus process is shown in Fig.~\ref{3}.

The consensus process of the T-RBFT algorithm is as follows:

1) Client request stage

The client sends request $ < {\rm{REQUEST}},op,seq,client{ > _{{\sigma _c}}}$ to the primary(that is, the primary ${s_i}$ of the consensus cluster) to start the inter-group consensus. Among them, $op$ represents the specific operation, and $seq$ is the request identifier: (1) The replicas (leaders of each group) store the $seq$ of the last request they executed for each client in the vector $Vreq$; (2) The replicas discard requests with $seq$ lower than the previous one (To avoid executing the same request twice), as well as discard any requests received while processing the previous request.

2) Inter-group consensus stage

The inter-group consensus steps are given in Algorithm~\ref{alg1}. The algorithm follows a message exchange pattern similar to PBFT, however the addition of the USIG service makes the entire inter-group consensus process only have two communication steps, instead of three communication steps like PBFT.

\begin{algorithm}
	\caption{Inter-group consensus algorithm}
	\label{alg1}
	\begin{algorithmic}[1]
		\upon{ ${s_i}$ wants to multicast a block message $m$}
	%	\begin{ALC@g} % ???????
			\State {$U{I_i}$ = \textsf{createUI}($m$)}
			\State {send $ \langle \text{PRE-PREPARE},v,{s_i},m,U{I_i}\rangle$ to leaders of  other groups}%all replicas
			%	\ENDWHILE
	%	\end{ALC@g}
		\upon{${s_j}$ receives $  \langle \text{PRE-PREPARE},v,{s_i},m,U{I_i} \rangle$  from ${s_i}$}	
	%	\begin{ALC@g} % ???????
			\If{\textsf{checkUI}$(P{K_i}, U{I_i}, m) $}
			\State	{ $U{I_j}$ = \textsf{createUI}($m$)}
			\State	{send $\langle\text{PREPARE },v,{s_j},{s_i},m,U{I_i},U{I_j}\rangle$ to leaders of  other groups}  %replicas
			\EndIf
			%	\ENDWHILE
	%	\end{ALC@g}
		\upon{${s_k}$ receives $\langle\text{PREPARE },v,{s_j},{s_i},m,U{I_i},U{I_j}\rangle $}
	%	\begin{ALC@g} 
			\If{${s_k}$ did not receive the PRE-PREPARE message and \textsf{checkUI}(${P{K_i},U{I_i},m}$) and \textsf{checkUI}(${P{K_j},U{I_j},m}$)}
			\State{$U{I_k}$ = \textsf{createUI}($m$)}
			\State{send $\langle\text{PREPARE },v,{s_k},{s_i},m,U{I_i},U{I_k}\rangle$ to leaders of  other groups}  %replicas
			\EndIf
			%	\ENDWHILE
	%	\end{ALC@g}
		\upon{${s_l}$ receives $\langle\text{PREPARE},v,{s_j},{s_i},m,U{I_i},U{I_j}\rangle$ from  $f+1$ different  group leaders for which \textsf{checkUI}(${P{K_j},U{I_j},m}$)}
	%	\begin{ALC@g} 

	 \State{enter intra-group consensus stage}

			%	\ENDWHILE
	%	\end{ALC@g}
	\end{algorithmic}
\end{algorithm}

When the primary ${s_i}$ wants to broadcast a request, ${s_i}$ sends a PRE-PREPARE message to the leader of each group (line 3), where $v$ represents the view number, $m$ is the request message sent by the client, and $U{I_i}$ is the unique identifier assigned by ${s_i}$ for the request.

Replica ${s_j}$ will perform the following four checks on the received PRE-PREPARE messages (lines 5 and 6): (1) The replica is currently in view $v$ and the PRE-PREPARE message is sent by the primary ${s_i}$; (2) The request contains a valid signature generated by client (to prevent a faulty primary from forging the request); (3) Check in $Vreq$ that the request has not been executed; (4) ${s_j}$ has received the request $m^{'}$ with $cv^{'}=cv-1$, where $cv$ is the counter value contained in the UI. The last condition supplements the unique identifier by enforcing FIFO order to prevent duplication, this is because unique identifiers by themselves do not protect against duplicity, and a faulty primary can still send inconsistent messages to different servers, even with different identifiers. By enforcing FIFO order on messages with a unique identifier ($UI$), all correct replicas process messages sent by the faulty primary in the same order (i.e. There will not be a correct replica processing the message $ < ...,{s_i},...,U{I_i},... > $ with a counter value of $cv$ sent by ${s_i}$ before processing the message $ < ...,{s_i},...,U{I_i}^{'},... > $ with a counter value of $cv - 1$ sent by ${s_i}$), or if one of the messages is not received, the correct replica will stop processing these messages to prevent duplication. Therefore, to enforce this property, each replica retains  a vector $Vacc$, which has the highest counter value $cv$ received from other replicas.

After passing these checks, ${s_j}$ broadcasts the PRE-PREPARE message to all nodes participating in the inter-group consensus (line 8), where $U{I_j}$ is obtained by calling createUI. Each message sent, whether PRE-PREPARE or PREPARE, has a $UI$ obtained by calling the createUI function (lines 2, 7, 13), so no two messages can have the same identifier.At the same time, this replica will receive PREPARE messages from other replicas. The replica determines whether to accept the PREPARE message by checking the message signature, view number, and using the checkUI function to check whether the received $UI$ is valid for $m$ and it appears in FIFO order.

If a replica ${s_k}$ does not receive a PRE-PREPARE message but receives a PREPARE message from the sender with a unique identifier $UI$, it sends its PREPARE message (lines 11-14). The purpose of doing this is to deal with two situations: (1) the sender  is faulty and does not send PRE-PREPARE message to ${s_k}$ (but to other replicas); (2) The PRE-PREPARE message was delayed and was received after the  PREPARE message.

When the replica receives valid PREPARE messages from $f+1$ different nodes for request $m$, the inter-group consensus is completed and enters the intra-group consensus stage (lines 17 and 18).

3) Intra-group consensus stage

At this time, the nodes participating in the inter-group consensus will serve as the leader of  group and execute the intra-group consensus.
The intra-group consensus steps are given in Algorithm~\ref{alg2}.

\begin{algorithm}
	\caption{Intra-group consensus algorithm}
	\label{alg2}
	
	\begin{algorithmic}[1]
		\State \textit{$\triangleright $ Leader  broadcast phase}
		\upon{leader $s_i$  receives  $f+1$ valid PREPARE messages  for  $m$ at term $T$ }
	%	\begin{ALC@g} % ???????
			\State{append ($T$, $m$) to local log $(logIndex = L$)}
			\State{generate leader's partial signature:\\
				$\sigma_i \gets \text{BLS\_Sign}(\text{sk}_i, H(T \mathbin\Vert L \mathbin\Vert m))$}
			\State $proofs \gets$ collect $f+1$ valid $UI$ proofs for $m$
			\For{each follower $r_j$}
			\State  send $\langle \text{AppendEntriesRPC}, prevLogTerm, 
			 prevLogIndex, leaderCommit,  $
			$ proofs,m,T,L,\sigma_i \rangle$ 
			\EndFor 
	%	\end{ALC@g} % ???????
		\State \textit{$\triangleright$ Follower reply phase}
		\upon{follower $r_j$ receives a AppendEntriesRPC meaasage from $s_i$}
	%	\begin{ALC@g} 
			\If{$T$ and $\sigma_i$ are valid}
			\If{\( \forall~UI_{i} \in proofs,\ \textsf{checkUI}(P{K_i},U{I_i}, m) \)}
			\State  $\sigma_j \gets \text{BLS\_Sign}(\text{sk}_j, H(T \mathbin\Vert L \mathbin\Vert m))$
			\State reply $\langle  \text{AppendEntriesReply}, true, \sigma_j \rangle$ to $s_i$
			\EndIf
			\EndIf
	%	\end{ALC@g} 
		\State \textit{$\triangleright$ Leader commit phase}
		\upon{$s_i$ receives  valid AppendEntriesReply meaasages more than $3n/4$ } 
	%	\begin{ALC@g} 
			\State	$\hat{\sigma_1} \gets \textsf{aggregate}(\{\sigma_j\})$ 
			\State apply  $m$ to state machine
			\State	broadcast $\langle\text{AppendEntriesCommit},T, L, m,{\hat{\sigma_1}\rangle}$ to all followers
	%	\end{ALC@g} 
		\State \textit{$\triangleright$ Commit reply phase}
		\upon{$r_j$ receives AppendEntriesCommit from $s_i$}
	%	\begin{ALC@g} 	
			\If{$\hat{\sigma_{1}}$ is valid and CountSigners$(\hat{\sigma_{1}}) \geq 3n/4$ }
			\State commitIndex $\gets L$
			\State apply  $m$ to state machine
			\State $\sigma^{'}_j \gets \text{BLS\_Sign}(\text{sk}_j, H(\text{"ACK"} \|T \mathbin\Vert L \mathbin\Vert m))$
			\State reply $\langle \text{AppendEntriesCommitReply},T,L,m,\sigma^{'}_j\rangle$ to leader $s_i$	
			\EndIf
	%	\end{ALC@g} 
	\end{algorithmic}
\end{algorithm}

\ding{172} Leader broadcast

When the leader ${s_i}$ receives  $f+1$ valid PREPARE messages  for  $m$ at term $T$, it first assigns the $m$ a unique log index $L$ and appends the tuple to its local log (Line 3). Then, the leader  generates a partial BLS signature $\sigma_i$ over the hash of the term, log index, and  block,  $H(T \mathbin\Vert L \mathbin\Vert m)$, using its private key ${sk}_i$ (Line 4). The leader ${s_i}$ broadcasts the AppendEntires  Remote Procedure Call (RPC) to all followers within the group. The AppendEntries RPC contains not only $m$, its current commit index, leaderCommit, and the signature $\sigma_i$ but also the $f+1$ $UI$ proofs obtained by  the inter-group consensus (Lines 5-7).

\ding{173} Follower reply

Upon receiving an AppendEntries RPC from the leader, each follower performs the following checks:
First, it verifies that the leader's term $T$ is not stale and that its local log matches the leader's claimed previous entry.
Then, it validates the leader's signature $\sigma_i$ against the hash $H(T \mathbin\Vert L \mathbin\Vert m)$ (Line 11).
Finally, it checks whether the attached sequence number in $UI$ matches its locally expected value, and whether the accompanying $f+1$ MACs over $H(m)$ are valid (Line 12). If all checks pass, the follower $r_{j}$ appends the entry to its log and generates its own partial signature $\sigma_j$ over the same hash $H(T \mathbin\Vert L \mathbin\Vert m)$, and replies to the leader with an AppendEntriesReply RPC message (Lines 13-14). 

\ding{174} Leader commit

Once the leader collects successful AppendEntriesReply RPC messages from a quorum of followers, it aggregates these signatures with its own $\sigma_i$  into a single threshold signature (Line 19). Then, the leader  commits the entry locally (updating commitIndex) and  broadcasts an AppendEntriesCommit RPC message  to  followers (Lines 20-21).

\ding{175} Commit reply

When the follower receives the AppendEntriesCommit RPC message sent by the leader, it verifies whether the aggregated signature $\hat{\sigma_1}$ is valid  and whether it contains at least $3n/4$ signatures. If the verification succeeds and the log index $L$ exceeds the locally recorded commitIndex, the node performs the following operations: it updates its local commitIndex to $L$ (Line 25), applies $m$ to the state machine (Line 26), generates a partial signature $\sigma^{'}_j$ for the acknowledgment message $H(\text{"ACK"} | T \mathbin\Vert L \mathbin\Vert m)$ (Line 27), and finally replies to the leader with an AppendEntriesCommitReply RPC message (Line 28).

4) Reply stage

The leader waits for AppendEntriesCommitReply RPC messages from at least $3n/4$ followers, each of which indicates that the block has been successfully committed to their local state. Upon collecting these confirmations, the leader aggregates the returned signatures into a new aggregated  signature, denoted as $\hat{\sigma_2}$. Then, the leader of each group sends a response message including $\hat{\sigma_2}$ to the client. If the client receives messages from more than $f +1$ groups, the consistency consensus process is completed.

\subsection{Leader election}
\subsubsection{View change in inter-group consensus}
When each replica receives a request, a timer will be started. If the request is still not executed after the timer times out, the replica will believe that the primary is faulty and  trigger the VIEW-CHANGE mechanism. At this time, this replica will send a VIEW-CHANGE message $ <{\rm{VIEW-CHANGE}},{s_i},v + 1,{C_{latest}},O,U{I_i}> $ to other replicas. Among them, ${C_{latest}}$ is a set representing the proof of generating the latest stable checkpoint (including $f + 1$ valid checkpoint information). $O$ is a collection of all messages sent by the replica since the latest checkpoint was generated (PRE-PREPARE, PREPARE, VIEW-CHANGE and NEW-VIEW messages).

The correct replica will perform the following three checks on the received VIEW-CHANGE message: (1) ${C_{latest}}$ contains $f + 1$ valid $UI$ identifiers; (2) The counter value in $U{I_i}$ is $cv+1$, where $cv$ is the highest counter value of the $UI$s signed by the replica in $O$; If $O$ is empty, the highest counter value will be the $UI$ in ${C_{latest}}$ when the replica generates the checkpoint; (3) The message sequence numbers in $O$ are continuous (no holes).

After the primary of the new view $v+1$ receives $f + 1$ valid VIEW-CHANGE messages, it broadcasts the NEW-VIEW message $ <{\rm{NEW-VIEW}},p,v + 1,{V_{cc}},S,U{I_p}> $ to all other nodes, where ${V_{cc}}$ is a set, including $ f + 1$ (including the new primary itself) VIEW-CHANGE messages, and $S$ is a set of requests prepared/accepted since the checkpoint. To calculate $S$, the primary first finds the latest stable checkpoint from all VIEW-CHANGE messages, and then selects requests from the $O$ set in the VIEW-CHANGE message, whose $UI$ counter value is greater than those in the checkpoint certificate.

After receiving the NEW-VIEW message broadcast by the primary, other replicas verify the signature of the message, calculate $O$ in the same way, and check whether it is the same as the $O$ calculated by the primary. If it is the same, the replica accepts the NEW-VIEW message.

Because the sequence number is provided by different tamper-resistant components (or USIG service) for each view, the replica will only start a new view $v+1$ after executing all requests in $S$ that were not previously executed. In addition, when the view is changed, the value of the first sequence number for the new view should be the counter value in $U{I_i}$ in the NEW-VIEW message plus one. The new primary send the next PRE-PREPARE message should follow the $U{I_i}$ in the NEW-VIEW message.

\subsubsection{Leader election in intra-group consensus}
If there are Byzantine nodes, the leader election mechanism of  Raft algorithm cannot continue to ensure the safety of the algorithm. Therefore, the intra-group consensus algorithm designs a new leader election mechanism (named Committed Proof) based on Raft, and additionally requires that the leader must support the trusted execution environment. The steps for leader election are as follows:

\ding{172} When the follower does not receive the heartbeat message sent by the leader before the heartbeat timer expires, it determines that the current leader is no longer working and is ready to initiate a leader election. Before the election, this follower first determines whether it has a TEE. If it has, it will  increase the term number by one, then change its  status to a candidate. Subsequently, it votes for itself and sends a RequestVote RPC to other followers in the cluster to request them to vote; If this follower does not have a TEE, its heartbeat timeout timer will be restarted.

\ding{173} When other followers receive the RequestVote RPC request sent by  candidate, the follower first determines whether the candidate's term is higher than itself or whether the log entry corresponding to the candidate is at least as new as its own. If not, it responds with a reject voting message. Otherwise,  it sends the term and index of its last consensus reached log entry as a response message parameter to candidate, and  initiates a remote attestation challenge to candidate. The purpose of the second phase is to require candidate to prove to the current follower that candidate's log list contains the last committed log entry of the current follower, that is, candidate needs to prove that it has a more complete log list than the voting node (committed proof).

\ding{174} After candidate receives the request message from follower, if it does have a log entry with the same index and term, it will return a hash value of the log as a response parameter. In addition,  it is necessary to send identity attestation (remote attestation function of the trusted platform) to the followers.

\ding{175} When the follower receives the response message from candidate, it verifies whether the received hash value is consistent with the hash value of its own corresponding log, and whether the remote attestation signature and integrity log of candidate are valid. %(i.e.  whether the candidate has a TEE). 
If the verification is passed, the vote is confirmed, otherwise the vote is rejected.

\subsection{Handling of Cross-Group Transactions}
Although T-RBFT adopts a hierarchical and multi-group structure, it inherently solves the challenge of cross-group transaction consistency. Specifically, T-RBFT adopts a two-layer consensus mechanism, in which the improved Raft algorithm is used for intra-group consensus and the BFT algorithm based on the trusted execution environment is used for inter-group consensus.

Therefore, transactions that involve multiple groups are coordinated by the BFT-based inter-group consensus, ensuring system-wide consistency and atomicity without the need for additional cross-group transaction protocols.

\section{Analysis and Proof of T-RBFT}
This section sketches proofs of the correctness of T-RBFT. We have to prove that the safety property is always satisfied (i.e., that all non-faulty replicas execute the same requests in the same order) and the same for liveness (i.e., that all clients' requests are eventually executed).
\subsection{Safety}
\subsubsection{Inter-group consensus}
\textbf{Normal case operation:} Since the entire protocol carries a signature, the primary cannot create a request, otherwise it will be discovered by honest nodes. Secondly, since the USIG service ensures that two different messages never get the same $UI$ identifier and therefore will not get the same sequence number, it ensures that a faulty primary cannot assign the same sequence number to two different request. Additionally, each replica stores the $seq$ of the last request executed for each client in a vector $Vreq$. Therefore, the replica will not execute the sequence number of the request that already exists in  $Vreq$, thus avoiding that when primary is faulty, it can order the same request twice.

\textbf{View change:} In the PBFT consensus algorithm, the commit phase ensures that the submitted request will be replayed in the new view, thereby solving the cross-view consistency problem. In the inter-group consensus of this article, we need to prove that the submitted request will also be replayed in the new view.

Assume that the ${V_{cc}}$ of NEW-VIEW contains a  message $<{\rm{VIEW-CHANGE}},r,{v^{'}},{C_{latest}},O,U{I_r}>$ for replica $r$  ($r\in Quorum$), we consider the following four situations:

1) The primary $p$ is correct and there is a correct replica $r$ that executed the request message $m$:

\ding{172} $m$ is executed after the stable checkpoint: since $r$ is correct, the VIEW-CHANGE message of $r$ must contain the PREPARE message sent by $r$ for $m$.

\ding{173} $m$ is executed before the stable checkpoint: this message will not need to be replayed in the new view.

2) The primary $p$ is correct but there is no correct replica in quorum $Q$ to execute request $m$: In this case, there must be a faulty replica $r \in Quorum$ that sent a PREPARE message about $m$. This is because a correct replica $s$ executed a request message $m$ with sequence number $i$ in view $v$, so it must have received $f + 1$ valid PREPARE messages from a quorum $Q^{'}$ of $f + 1$ replicas. Due to $|Q|+|Q'|=2f+2$. Therefore, there must be a replica that sent a PREPARE message about $m$.

\ding{172} $m$ is executed after the stable checkpoint: $r$ may not put $m$'s PREPARE message into $O$, and if it does, then $p$ will not put $r$'s VIEW-CHANGE message into ${V_{cc}}$. The reason is that If $r$ wants not to put the PREPARE message of $m$ into $O$, it needs to do one of two things, but this will be detected by $p$:

a) If $r$ executes another request $m^{'}$ after $m$, then $r$ may put the PREPARE message of $m^{'}$  into $O$ instead of the PREPARE message of $m$, which will leave a "hole" in $O$ that is detected by $p$.

b) If $r$ assigns $UI$ with counter value $cv$ to $m$'s PREPARE messages, $r$ may not put all PREPARE messages of $cv^{'} \ge cv$ into $O$, but this will be detected by $p$, because $r$ has to put the VIEW-CHANGE messages of $cv^{''} > cv + 1$ into $O$. Therefore, if $r$ wants $p$ to put the VIEW-CHANGE message of $r$ into ${V_{cc}}$ , then $r$ needs to include the PREPARE message of $m$ in $O$ of the VIEW-CHANGE message it sends.

\ding{173} $m$ is executed before the stable checkpoint: A faulty replica $r$ may put an old checkpoint into the VIEW-CHANGE message, but similarly, $p$ will not put $r$'s VIEW-CHANGE message into ${V_{cc}}$ for the same reason as \ding{172}.

3) The primary $p$ is faulty but there is a correct replica $r$ in quorum $Q$ to execute request $m$: In this case, the faulty $p$ can try to modify the content in ${V_{cc}}$. For example, it can delete $m$ or replace the checkpoint certificate ${C_{latest}}$. But this can all be detected, because the correct replica node will verify its validity after receiving the NEW-VIEW message from the primary $p$. However, $p$ cannot forge the $UI$ of $r$. Therefore, a faulty primary cannot tamper with the contents of  correct replica's VIEW-CHANGE message.

4) The primary $p$ is faulty and there is no correct replica $r$ in quorum $Q$ to execute request $m$: In this case, there must be a faulty replica $r \in Quorum$ that sent a PREPARE message about $m$. In case 2, we have proven that $r$ can not make $p$ believe that it did not execute $m$. However, in case 4, $p$ is also faulty, so the faulty $p$ can still try to modify the VIEW-CHANGE message sent by $r$ in ${V_{cc}}$. However this belongs to case 3, because the correct replica node will verify its validity after receiving $p$'s NEW-VIEW message.
\subsubsection{Intra-group consensus}
In the intra-group consensus, followers are able to independently verify message proofs before accepting blocks by leveraging a trusted execution environment and verifiable aggregate signatures. This design ensures the security of the system in the presence of Byzantine nodes, including the security of the situations discussed in Section 2.2:

1) Log replication

\ding{172} Assume that the leader is a Byzantine node:
To defend against a Byzantine leader that may tamper with transactions or send inconsistent blocks to different followers, our design incorporates authenticated inter-group consensus proofs and local verification mechanisms within each group. After inter-group consensus, the leader of a group obtains a certified certificate, which is a set of $f+1$ $UI$ identifiers from distinct group leaders on the block meassage. This certificate is attached to the block and disseminated to followers within the group during the AppendEntries phase.

Upon receiving the block, each follower verifies the attached certified certificate (e.g., by checking the  $f+1$ MACs) to ensure that the he block has been approved by a sufficient number of group leaders in the inter-group consensus and the block content (its hash) matches what was agreed upon.

In addition, each block includes a strictly increasing sequence number assigned during inter-group consensus, which allows followers in the group to detect any attempt by a byzantine leader to reorder or revert blocks.

Thus, even if the leader is a Byzantine node, it cannot forge a valid authentication certificate nor send different blocks to different followers without detection. This prevents inconsistent log replication and ensures that only globally agreed blocks are committed across the system.

\ding{173} Assume that the follower is a Byzantine node: Assuming that each group contains $n$ nodes with $f$ Byzantine follower nodes, and the Byzantine follower nodes may tamper with the received block content, add it to their own blockchain, and reply to the leader that they have successfully uploaded the block to the chain. Therefore, out of  $x$ nodes reply received by the leader, there may be $f$ from Byzantine nodes. To ensure that the AppendEntries messages are accurately sent to the majority of honest nodes, $x$ should satisfy: $x - f > n/2$. As can be seen from the previous text, $f < n/4$ can be obtained when leader  receives responses from more than $3n/4$ nodes. Therefore, intra-group consensus can ensure the normal progress of consensus when the number of Byzantine nodes in the group is less than $n/4$.

2) Leader election

During the leader election phase, a candidate needs to prove that it has a more complete log list than other nodes through secure hash values, thereby preventing the Byzantine candidate from setting a "newer" logical timestamp than actual in the RequestVote message to obtain votes from other nodes and reducing the possibility of Byzantine nodes being elected as leader. In addition, the remote attestation technology of trusted computing ensures that the new leader has a trusted execution platform, limiting its malicious behavior during the consensus process.

3) Consensus confirmation

\ding{172} Assume that the leader is a Byzantine node:
After the leader receives more than $3n/4$ of signed AppendEntriesCommitReply RPC messages from followers, it considers that the block has reached consensus within the group and aggregates these signatures. Subsequently, the leader returns the aggregated signature along with the response to the client as proof that the block has indeed reached consensus. Therefore, if not enough AppendEntriesCommitReply messages are received, the leader cannot forge a group-level commit proof and return it to the client.

Thus, the client will only receive confirmation when the block has truly been committed by a quorum of nodes in the group, ensuring that the leader cannot lie about consensus status.

\ding{173} Assume that the follower is a Byzantine node: Since the leader  has received replies from more than $3n/4$ nodes, it is guaranteed to have received replies from more than half of  honest followers. Additionally, since the log list of this follower will lack log entries that have already reached consensus in the cluster,  leader election mechanism can ensure that this follower no longer has the ability to be elected as leader, thus ensuring the safety of the consensus algorithm.
\subsection{Liveness}

The liveness of this algorithm is provided by the view change during inter-group consensus stage and leader election during  intra-group consensus stage. In the inter-group consensus stage, replicas determine the primary through $p = v\bmod k$, and the view number v increases from 0. When the timer of the replica expires or receives $f + 1$ valid view change messages, it means that the primary of current view is no longer able to complete consensus, and replicas need enter the next view to resume consensus service. In addition, the node conducting inter-group consensus will also serve as the leader of the intra-group consensus, and the followers within the group will keep in touch with  leader through heartbeat messages. If there is no response within the timeout, the leader election protocol will be triggered for leader re-election, and the new leader will replace the previous leader to participate in  inter-group consensus, which accelerates the detection and replacement of faulty nodes by  inter-group consensus, and enhancing the liveness of  inter-group consensus.

During the intra-group consensus stage, the leader election protocol and heartbeat detection mechanism ensure the liveness of  intra-group consensus algorithm. When the follower fails to receive the leader's heartbeat message within the timeout period, it will transition to candidate status for new leader election, thereby ensuring the continuation of consensus.

\subsection{Communication overhead analysis}

This subsection analyzes the communication times required for a single consensus of the T-RBFT algorithm. For the convenience of analysis, assuming that the consensus nodes in the blockchain are divided into $k$ groups and the number of nodes in each group is $n$. The total number of nodes in the system is
$N = k*n$.

Our consensus process is divided into inter-group consensus and intra-group consensus. the number of communication times for inter-group consensus is $k^{2}-1$. For intra of a single group,  leader first broadcasts the block to all followers. The number of communication times in this process is $n-1$; Then, followers will reply to leader and the number of communication times in this process is $n-1$. Finally, the totla number of communication times of leader commit and commit reply is $2n-2$. Therefore, the number of communication times required for T-RBFT  to reach a consensus  can be obtained:
\begin{equation}
T=k^{2}-1+k*(4n-4)=k^{2}+4N-4k-1
\label{eq9}
\end{equation}

It can be seen from the analysis that compared with PBFT which has $O(N^{2})$ communication complexity, the communication complexity of T-RBFT is only $O(N + K^{2})$.

\section{Evaluation}
The experimental platform was a host server configured with an Intel(R) Core(TM) i7-9700 processor (8 cores) and 36 GB of RAM. A multi-node consensus environment was emulated on this host through container-based deployment combined with port mapping.  The software environment was Ubuntu 20.04 and Golang 1.16.

Due to the need for  trusted message authentication services, we borrowed the virtual counter method proposed in reference \cite{18}, which uses Intel SGX to provide hardware security support and implements the TEE part of  replica as  an SGX enclave. We use HMAC-SHA256 in the USIG service for message authentication, and the BLS signature algorithm for aggregating signatures in the intra-group consensus process. Currently, we assume that the key is  manually installed before system startup to implement the USIG service we described in Section 2.4.

\subsection{Communication times}
First, We compare the communication times required for a single consensus of the T-RBFT algorithm with the MinBFT \cite{16},  R-PBFT \cite{20} and WRBFT \cite{32}. Among them, MinBFT is independent TEE-based solutions, while R-PBFT and WRBFT are the latest consensus solutions that adopt the similar two-layer structure as proposed in this paper. We set the same number of nodes in each group of R-PBFT, WRBFT and our T-RBFT, where the number of nodes in each group is 4.

\begin{figure}[htbp]\centering
	\includegraphics[width=0.5\columnwidth]{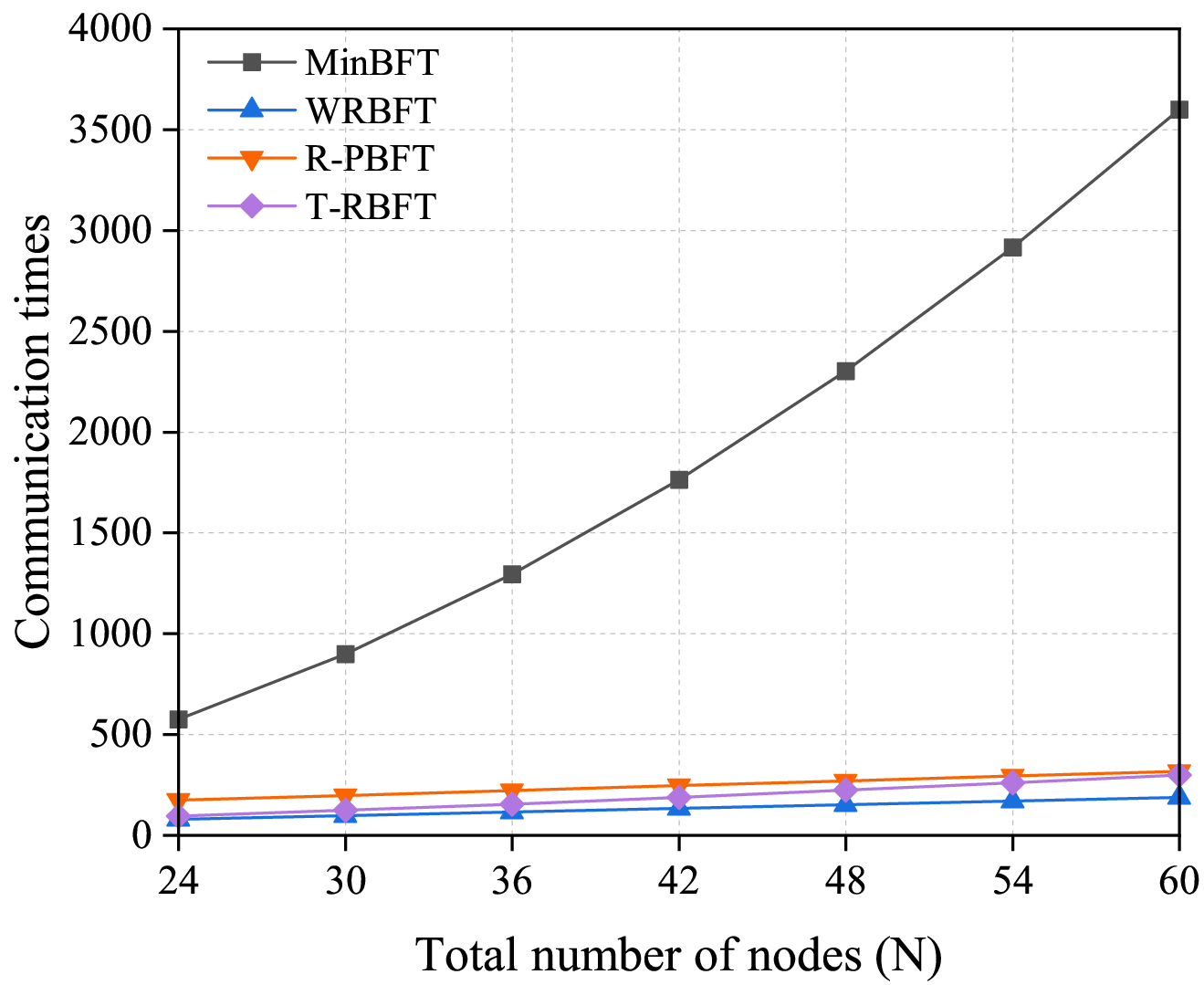}
	\caption{Comparison of communication times\label{4}}
\end{figure}

Fig.~\ref{4} shows the comparison of the number of communication times  for different algorithms to complete a consensus under different nodes. 
Under the same number of nodes, the WRBFT algorithm requires the least number of communication times, while MinBFT requires the most. The number of communication times required by our T-RBFT is slightly lower than that of  R-PBFT. When the number of nodes is 42, R-PBFT requires 246 communications to reach a consensus, while T-RBFT requires 188.

Table~\ref{tab1} shows communication times of T-RBFT when the number of nodes is 60. 
As the number of groups decreases, the number of communication times required for  T-RBFT algorithm to complete a consensus decrease. When the number of groups is 3, the minimum number of communication times can reach 236.
\begin{table}[htbp] % ????????????
	\belowrulesep=0pt\aboverulesep=0pt
	\centering % ??????
	\caption{Communication times of T-RBFT when the number of nodes is 60} % ????
	\label{tab1} % ?????????
	\begin{tabular}{>{\centering\arraybackslash}p{5cm} >{\centering\arraybackslash}p{5cm}}
		\toprule % ??????
		Number of groups & Communication times \\ % ????
		\midrule % ??????
		k=20 & 559 \\ % ???1
		k=15 & 404 \\ % ???2
		k=12 & 335 \\ % ???3
		k=10 & 299 \\ % ???4
		k=6  & 255 \\ % ???5
		k=3  & 236 \\ % ???6
		\bottomrule % ??????
	\end{tabular}
\end{table}

\subsection{Fault tolerance}
\begin{figure}[htbp]\centering
	\includegraphics[width=0.5\columnwidth]{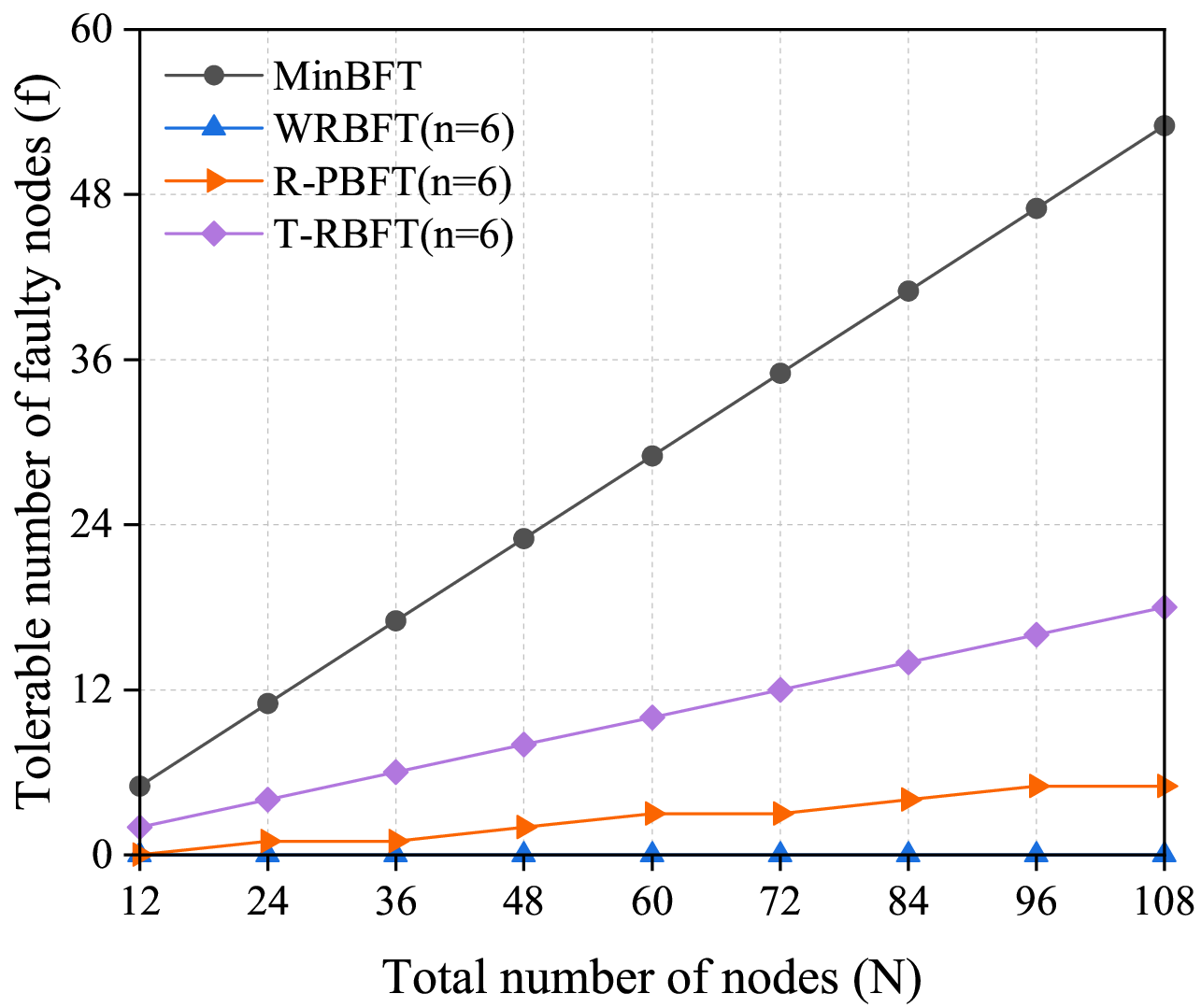}
	\caption{Comparison of fault tolerance under different number of nodes
		\label{8}}
\end{figure}

Fault tolerance is one of the important properties that affects the safety of the algorithm. Therefore, the fault tolerance performance of the T-RBFT consensus mechanism is compared with the  MinBFT, R-PBFT and WRBFT consensus mechanisms. The results are shown in Fig.~\ref{8}.

As can be seen from Fig.~\ref{8}, MinBFT achieves the highest fault tolerance, followed by T-RBFT, and then R-PBFT.  For example, when the network is divided into 6 groups, the number of nodes in each group is 6, so $N$=36. In this situation, the maximum number of  fault tolerance using MinBFT  is 17. The maximum number of fault tolerance  using T-RBFT is 6 (the number of fault tolerance for inter-group consensus is 2, and the number of fault tolerance within each group is 1), while using R-PBFT is only 1. In addition, we believe that the WRBFT protocol is obviously insecure, because in the inter-group consensus of WRBFT,  improving the original PBFT consensus only through BLS aggregate signatures cannot guarantee  its security. A faulty primary may assign the same sequence number to two different requests, but the replicas in the WRBFT  cannot detect this malicious behavior of the primary. Therefore, WRBFT cannot tolerate Byzantine nodes.

\section{Conclusion}
This paper presents T-RBFT,  a novel two-layer consensus mechanism suitable for consortium blockchain. Compared to  the existing two-layer consensus for consortium blockchain and BFT variants, T-RBFT achieves lower latency under the same fault tolerance threshold. Moreover, as the number of nodes increases, evaluation results show that T-RBFT exhibits better scalability with a more stable throughput, effectively addressing the performance bottlenecks of prior designs that rely on global, serial consensus.  Overall, T-RBFT can provide a robust, scalable, and efficient consensus foundation for permissioned blockchain networks, particularly suitable for concurrency scenarios and application domains that demand both trust and performance.  In the future, we plan to implement our solution on multiple standard TEE platforms (such as GlobalPlatform \cite{36}), and combine it with cross-chain technology to further solve the problems of low scalability and poor privacy isolation of single chain of consortium blockchain.

%\begin{itemize}
%\item Funding
%\item Conflict of interest/Competing interests (check journal-specific guidelines for which heading to use)
%\item Ethics approval and consent to participate
%\item Consent for publication
%\item Data availability 
%\item Materials availability
%\item Code availability 
%\item Author contribution
%\end{itemize}

%\section*{Declarations}

%Some journals require declarations to be submitted in a standardised format. Please check the Instructions for Authors of the journal to which you are submitting to see if you need to complete this section. If yes, your manuscript must contain the following sections under the heading `Declarations':

%\begin{itemize}
%\item Funding
%\item Conflict of interest/Competing interests (check journal-specific guidelines for which heading to use)
%\item Ethics approval and consent to participate
%\item Consent for publication
%\item Data availability 
%\item Materials availability
%\item Code availability 
%\item Author contribution
%\end{itemize}

\bibliography{sn-bibliography}% common bib file
%% if required, the content of .bbl file can be included here once bbl is generated
%%\input sn-article.bbl

\end{document}